\documentclass[aps,amsmath,amssymb,groupedaddress,]{article}
\usepackage{authblk}

\usepackage{geometry}
\usepackage{amsmath}
\usepackage{amsfonts}
\usepackage{amsthm}

\usepackage{graphicx}

\usepackage{amssymb}

\usepackage{booktabs}
\usepackage{caption} % didascalia Tabella
\def\unit#1{\ensuremath{\mathrm{\,#1}}}

\usepackage{xcolor} %pacchetto per poter evidenziare testo 

\title{Volume fraction determination of microgel composed of interpenetrating polymer networks of PNIPAM and polyacrylic acid}

\author[1,2]{Silvia Franco}
\author[2]{Elena Buratti}
\author[2,3]{Barbara Ruzicka}
\author[4]{Valentina Nigro}
\author[5]{Nicole Zoratto}
\author[5]{Paolo Matricardi}
\author[2,3]{Emanuela Zaccarelli}
\author[2,3]{Roberta Angelini}

\affil[1]{Dipartimento di Scienze di Base e Applicate per l'Ingegneria (SBAI), Sapienza Università di Roma, 00185 Roma, Italy;}
\affil[2]{Istituto dei Sistemi Complessi del Consiglio Nazionale delle Ricerche (ISC-CNR), Sede Sapienza, 00185 Roma, Italy;}
\affil[3]{Dipartimento di Fisica, Sapienza Università di Roma, 00185 Roma, Italy;}
\affil[4]{ENEA C.R. Frascati, FSN-TECFIS-MNF Photonics Micro and Nanostructures Laboratory, 00044 Frascati, Rome, Italy;}
\affil[5]{Dipartimento di Chimica e Tecnologia Farmaceutiche, Sapienza Universit\`a di Roma, 00185 Roma, Italy;}

\date{October 2020}

\begin{document}

\maketitle

\begin{abstract}

Interpenetrated polymer network microgels, composed of  crosslinked networks of poly(N-isopropylacrylamide) and polyacrylic acid (PAAc), have been investigated through rheological measurements at four different amounts of polyacrylic acid.
Both PAAc content and crosslinking degree modify particle dimensions, mass and softness, thereby strongly affecting the volume fraction and the system viscosity.
Here the volume fraction is derived from the flow curves at low concentrations by fitting the zero-shear viscosity with the Einstein-Batchelor equation which provides a parameter $k$ to shift weight concentration to volume fraction.
We find that particles with higher PAAc content and crosslinker are characterized by a greater value of  $k$ and therefore by larger volume fractions when compared to softer particles.
The packing fractions obtained from rheological measurements are compared with those from static light scattering  for two PAAc contents revealing a good agreement.
Moreover, the behaviour of the viscosity  as a function of packing fraction, at room temperature, has highlighted an Arrhenius dependence for microgels synthesized with low PAAc content and a Vogel-Fulcher-Tammann dependence for the highest investigated PAAc concentration.
A comparison with the hard spheres behaviour indicates a steepest increase of the viscosity with decreasing particles softness. 
Finally, the volume fraction dependence of the viscosity at a fixed PAAc and at two different temperatures, below and above the volume phase transition, shows a quantitative agreement with the structural relaxation time measured through dynamic light scattering indicating that IPN microgels softness can
be tuned with PAAc and temperature and that, depending on particle softness, two different routes are followed.
\end{abstract}

\section{Introduction}
Soft colloids have recently attracted great attention both in fundamental studies and applications thanks to the peculiarity of their particles to be deformable, elastic and potentially interpenetrable, providing a very rich phenomenology  \cite{MattssonNature2009, LikosSM2006, BerthierRMP2011, Brijitta2019, GnanNature2019, NigroJCIS2019}.
At variance with the largely studied hard sphere systems, soft colloids can display behaviours more intriguing  since they are characterized by many fascinating properties: the most remarkable is their softness, that depends on the underlying macromolecular architecture (microgels, star polymers, core-shell particles, liposomes...) and can also be tuned during the synthesis process \cite{LikosSM2006, VlassopoulosCOCIS2014}.
Softness makes these systems very versatile for technological applications so as the fabrication of new smart materials \cite{ DasARM2006, VinogradovCPD2006, BaglioniNanoscale2012, ParkBio2013, Karg2019, MazzucaACSAPM2020}.
Soft particles can be deformed and compressed until they reach non spherical shapes in order to be packed in all the available volume achieving high volume fraction values well above 1.
This allows to directly connect the degree of particle softness to the increasing volume fraction, which is the fraction of the total volume that is filled by N spheres, each of volume $n$. For hard-spheres, the volume fraction is defined as $\phi$=$nV$, where $n$ is the number density of particles. For microgels, $\phi$ is no longer a good measure of volume fraction since particles are deformable and their volume is not fixed. In this case we use the generalized volume fraction (sometimes called effective volume fraction) $\zeta=n V_{0}$, where $V_{0}=4/3 \pi R_{0}^{3}$ is the volume of a swollen particles of radius $R_{0}$ measured in dilute conditions. Therefore, a generalized volume fraction $\zeta$ is the total volume occupied by the swollen particles divided by the volume of the suspension \cite{MattssonNature2009, VanderScheerACSNano2012, RomeoSM2013, UrichSM2016, ScottiSM2020, BergmanNatComm2018}. At low concentrations, where particles volume does not depend on concentration, microgels behave as hard spheres and $\zeta=\phi$.
Volume fraction is an important parameter since its knowledge makes possible a direct comparison among experiments, simulations and theory.
One of the open problems for soft colloids is the univocal determination of their volume fraction.
In general, the experimental determination of volume fraction is not obvious since real colloids, even hard ones \cite{PoonSM2012, RoyallSM2012}, have always some degree of softness in their interparticle potential.
Moreover, they are polydisperse i.e. they have a finite size distribution \cite{PoonSM2012}.
For soft colloids this issue is even more complex due to changes in their volume and particle deformation.
%Therefore, a generalized volume fraction $\zeta$, the volume occupied by the swollen particles divided by the total volume of the suspension, is introduced \cite{MattssonNature2009, VanderScheerACSNano2012, RomeoSM2013, UrichSM2016, ScottiSM2020}. 
%
Among soft colloids, microgels, aqueous dispersions of nanometer or micrometer-sized particles composed of chemically crosslinked polymer networks, are very intriguing due to their hybrid nature which combines the properties of polymers and colloids \cite{Karg2019, LyonRevPC2012, StiegerLang2004, RovigattiSM2019}.
Microgel volume fraction depends on the crosslinking degree that controls particle softness but, in the case of responsive microgels, it also depends on external parameters (temperature, pH, etc...).
The experimental determination of volume fraction can be carried out in different ways: from rheological measurements through the viscosity curve in very dilute condition \cite{ScottiSM2020, SenffCPS2000, deKruifJCP1985},  from static light scattering through the Zimm plot \cite{Tuzar1996} and from small-angle neutron scattering using the zero average contrast method \cite{MohantySciRep2017, NojdSM2018}.
\\
Here we report an experimental study on the determination of the volume fraction of an interpenetrated polymer network (IPN)  microgel composed of poly(N-isopropylacrylamide) (PNIPAM) and polyacrylic acid (PAAc).
At variance with microgels obtained by random copolymerization (NIPAM-co-AAc), which are composed of a single network of both monomers, with properties dependent on the monomer ratio, microgels obtained by polymer interpenetration IPN, are made of two interpenetrated homopolymeric networks of PNIPAM and PAAc, each preserving its independent responsiveness to the external stimuli.
The former is a thermo-sensitive polymer characterized by a Volume Phase Transition (VPT) occurring  at about 305$\unit{K}$ that drives the system from a swollen hydrated state to a shrunken dehydrated one, as a consequence of the coil-to-globule transition of PNIPAM chains. The latter is a pH-sensitive polymer that shows a transition from hydrophobic to hydrophilic state above pH 4.5, corresponding to the pKa of acrylic acid.
The addition of a second network, physically interlaced with the first one, influences the particle softness, since the swelling capability is reduced due to the increase of topological constraints \cite{NigroJCIS2019}. Indeed, IPN microgels with higher PAAc content require more crosslinker reacting during synthesis leading to the formation of stiffer particles.
Previous studies on IPN microgels have investigated the behaviour of their radius across the VPT at different experimental conditions \cite{XiaLangmuir2004, NigroJCP2015, NigroCSA2017}, their morphology \cite{NigroJCIS2019, LiuPolymers2012} and local structure \cite{NigroJCP2015, NigroJML2019}, their behaviour in different solvents (H$_{2}$O and D$_{2}$O) \cite{NigroSM2017} and the role of softness \cite{MattssonNature2009, NigroJCIS2019, NigroMacromol2020}.
Moreover, they have been tested for drug release \cite{HuAdvMater2004, XiaJCR2005} and in vivo controlled release \cite{ZhouBio2008}.
%
%
%At variance with previous studies, in the present work we determine the volume fraction of IPN microgels composed of PNIPAM and PAAc through viscosity measurements.
%
%
%
%
At variance with previous studies on IPN, in the present work we determine the volume fraction of IPN microgels composed of PNIPAM and PAAc through viscosity measurements at different PAAc and crosslinker contents in order to understand the role of both parameters on particle softness and packing ability. We find that particles with higher PAAc and crosslinker content, which are therefore stiffer, are characterized by larger volume fractions, and that dynamics slows down in two different ways for low and high PAAc contents and above and below the VPT depending on particle softness. The manuscript is structured as follows:
firstly, experimental flow curves for microgels at four PAAc contents and different concentrations are shown and fitted through the Cross model to obtain the zero-shear viscosity.
Then, the relative zero-shear viscosity is reported as a function of concentration for different samples and compared to the Einstein-Batchelor equation in dilute regime.
A factor $k$ to convert the weight concentration in volume fraction is obtained.
\\
The volume fractions derived through viscosity are in good agreement, within the experimental error, with those measured through Static Light Scattering (SLS) \cite{MicaliCPC2018}.
Moreover, we report the viscosity behaviour as a function of volume fraction showing that it follows an Arrhenius dependence at low PAAc content and a Vogel-Fulcher-Tammann dependence for the highest investigated PAAc concentration.
These findings  indicate a steepest increase of the viscosity with increasing PAAc
content and therefore decreasing particles softness. Finally, an excellent agreement between viscosity and structural relaxation time at two temperatures, respectively below and above the VPT, is found corroborating the idea that softer particles display a dynamical slowing down at higher volume fractions when compared to stiffer ones.

\section{Determination of volume fraction from shear viscosity}

\subsection{\textit{Cross equation for the shear rate dependence of viscosity}}
Microgel suspensions at low concentration display a shear rate $(\dot{\gamma}$) dependent flow behaviour \cite{FernandezNievesWyss,SenffCPS2000} characterized by three different regions: two plateau regions at low and high shear rates ($\eta_{0}$ and $\eta_{\infty}$ respectively) and a shear thinning region at intermediate values.
The flow curves of microgels can be described through the Cross model \cite{Makosko}: 
\\
\\
\begin{equation}
\eta=\eta_{\infty}+\frac{\eta_{0} - \eta_{\infty}}{1+(\frac{\dot{\gamma}}{\dot{\gamma}}_{c})^{m}}
\label{cross_model}
\end{equation}
where $\eta_{0}$ and $\eta_{\infty}$ are the limiting viscosities at zero and infinite shear rate respectively, $\dot{\gamma}_{c}$ is an intermediate critical shear rate and $m$ is a positive power exponent.
\\
\subsection{Einstein-Batchelor equation for the concentration dependence of viscosity in dilute regime}
Rheological properties of diluted suspensions depend on the volume fraction $\phi$ (or packing fraction) of the dispersed components \cite{Bechinge2013}. In dilute regime this can be estimated by shear viscosity measurements.
Polymer chains, constituting microgel particles, contribute to increase the system viscosity linearly with polymer concentration beyond the viscosity value of the solvent $\eta_{s}$.
At low concentration the virial expansion of viscosity is \cite{PolymerPhysics,QuemadaRA1977}:
\begin{equation}
\eta=\eta_{s}(1+[\eta]C_w+k_{H} [\eta]^{2} C_w^{2}+...)
\label{eq1}
\end{equation}
where $[\eta]$ is the {\it intrinsic viscosity} that for microgels is related to the swelling capability in response to different solvents \cite{PolymerPhysics}, \textit{k}$_{H}$ is the Huggins coefficient representing the viscosity second virial coefficient and C$_w$ the weight concentration.
Equation (\ref{eq1}) can be rewritten as:
\begin{equation}
\frac{\eta-\eta_{s}}{\eta_{s} C_w}=[\eta]+k_{H} [\eta]^{2} C_w+...
\label{eq2}
\end{equation}
where $(\eta-\eta_{s})/\eta_{s}$ is the {\it relative viscosity increment} and $(\eta-\eta_{s})/\eta_{s}C_w$ the {\it reduced viscosity}.
Furthermore, $[\eta]$ can be expressed as the zero-concentration limit of the reduced viscosity:
\begin{equation}
[\eta]=\lim_{C_w \to 0} \frac{\eta_{0}-\eta_{s}}{C_w \eta_{s}}
\label{eq3}
\end{equation}
where $\eta_{0}$ is the {\it zero-shear viscosity}, $(\eta_{0}-\eta_{s})/\eta_{s}$ is a dimensionless quantity and the intrinsic viscosity [$\eta$] is proportional to the reciprocal of the concentration.
\\
For hard spheres the intrinsic viscosity $[\eta]$ is equal to 2.5 \cite{PolymerPhysics,Tadros} and the zero shear viscosity in this case can be written through the Einstein equation:
\begin{equation}
\eta_{0}-\eta_{s}=2.5 \eta_{s} \phi.
\label{eq4}
\end{equation}
Defining the {\it relative viscosity}: 
\begin{equation}
\eta_{rel}=\frac{\eta_{0}}{\eta_{s}} 
\label{etarel}
\end{equation}
equation (\ref{eq4}) becomes:
\begin{equation}
\eta_{rel}=1+2.5 \phi=1+[\eta] \phi .
\label{eq5}
\end{equation}
By replacing equation (\ref{eq4}) in (\ref{eq3}) the  volume fraction occupied by a colloidal particle, expressed in terms of concentration, is:
\begin{equation}
\phi=kC_w
\label{phi}
\end{equation}
that allows to connect weight concentration C$_{w}$ to  packing fraction $\phi$.
The volume fraction of colloidal suspensions in dilute conditions is estimated from the relative viscosity $\eta_{rel}$ through Einstein-Batchelor equation \cite{BatchelorJFM1977}:
%unire bibliografia [4-8]
%
%
\begin{equation}
\eta_{rel}=1+2.5 \phi +5.9 \phi^{2}
\label{EB}
\end{equation}
and replacing equation (\ref{phi}) in (\ref{EB}) one gets the relation of the relative viscosity of dilute suspensions as a function of weight concentration C$_{w}$:

\begin{equation}
\eta_{rel}=1+2.5 k C_w +5.9 k^2 C_w^{2}.
\label{EBCw}
\end{equation}
Therefore, fitting the relative viscosity of dilute suspensions as a function of weight concentration C$_{w}$ through equation (\ref{EBCw}),  the conversion constant $k$ between volume fraction $\phi$ and weight concentration C$_{w}$ can be obtained.
\\
Nevertheless, the parameter $\phi$ usually defines the volume fraction of hard sphere suspensions and it is expressed as $\phi$=$nV$, where $n$ is the number density and $V$ the volume of particles.
Instead, for soft particles, whose volume is not fixed because of their deformability, the parameter $\zeta$ is introduced. It is generally referred to a {\it generalized volume fraction} or sometimes as {\it effective volume fraction} and it is  defined as: 
\begin{equation}
\zeta = n V_{0}
\label{zetavolume}
\end{equation}
where $ V_{0}$ is the volume of an undeformed particle $V_{0} = \frac{4}{3} \pi R_{0}^3$
with $R_{0}$ the radius measured in dilute conditions. In these conditions soft particles can be treated as hard spheres and $\zeta=\phi$ \cite{MattssonNature2009, RomeoSM2013}. This assumption is not true out of this regime where, at variance with hard spheres, microgel volume can change and particles can deform and modify their size as a consequence of the variation of control parameters such as concentration, temperature and pH.
For this reason, in this work we use the generalized volume fraction $\zeta$  to describe the behaviour of soft IPN microgels \cite{MattssonNature2009, RomeoSM2013, UrichSM2016, ScottiSM2020} so that equation (\ref{phi}) becomes:
\begin{equation}
\zeta= k C_{w}
\label{costante_k}
\end{equation}
\subsection{\textit{Arrhenius and Vogel-Fulcher-Tamman models for the concentration dependence of viscosity}}
Outside the diluite regime a deviation from the Einstein-Batchelor equation is found, in fact viscosity of colloidal suspensions strongly depends on concentration and generally increases with it.
In order to describe this behaviour, several models are used but the widespread empirical growth is an Arrhenius-like dependence \cite{Bagley1983, Berthier2009} characterized by an exponential with two free parameters:
\begin{equation}
\eta=\eta_{0} \, exp \Big( A \zeta \Big)
\label{Arrhenius}
\end{equation}
where $\eta_{0}$ is the viscosity value in the limit of $\zeta$=0 and A controls the growth of the function.
This dependence marks a low sensitivity of viscosity to small changes in concentration.
Instead, if the viscosity is highly sensitive to changes in concentration, it is well described by the the Vogel-Fulcher-Tamman (VFT) model \cite{Philippe2018, ShamanaSM2018, SeekellSM2015} identified by an exponential with three free parameters:
\begin{equation}
\eta=\eta_{0} \, exp \Big( \frac{A \zeta}{\zeta_{0}-\zeta} \Big)
\label{VFT}
\end{equation}
where $\eta_{0}$ is the viscosity in the limit of $\zeta$=0, A is the growth parameter and $\zeta_{0}$ is the critical volume fraction that signs the divergence of $\eta$ \cite{ShamanaSM2018, VanDerScheerACSN2017}.
\\
This behaviour shares many analogies with molecular glasses approaching the glass transition temperature T$_{g}$ where viscosity increases by several orders of magnitude. In fact, according to the Angell classification \cite{AngellScience1995}, if viscosity is slowly sensitive to changes in  temperature, it is well described by an Arrhenius behaviour and the glass former can be classified as ‘‘strong’’ while if it has a much higher temperature sensitivity, it is well described by a super Arrhenius behaviour (Vogel-Fulcher-Tammann or power-law divergence) and the glass is defined ‘‘fragile’’.

\section{Determination of volume fraction from molecular weight} 
As described in section 2.2, microgel particles are deformable, their volume is not fixed and the generalized volume fraction  $\zeta$  of eq. \ref{zetavolume} is used, where the number density $n$ can be defined as:
\begin{equation}
n = \frac{N_{A} \rho}{M_{w}}
\label{n}\
\end{equation}
with $N_{A}$ is the Avogadro number, $\rho$ the mass density related to the weight concentration of the suspension and $M_{w}$ the molecuar weight.
\noindent
\\
Our aim is to compare the volume fraction obtained from viscosity measurements, $\zeta_{\eta}$, with the one obtained from the molecular weight, $\zeta_{M_w}$ derived from Static Light Scattering (SLS) measurements of a previuos work \cite{MicaliCPC2018}.
\\
SLS measurements on IPN samples were performed at various PAAc contents and  concentrations, in dilute regime ((5$\cdot$10$^{-6}$ - 5$\cdot$10$^{-5}$) g cm$^{-3}$) \cite{MicaliCPC2018}.
\\
The absolute excess scattered intensity $I(Q)$, i.e., the excess Rayleigh ratio R$_{ex}$ in cm$^{-1}$, was calculated from the measured scattered intensity profile $I^{meas}(Q)$ as:
\begin{equation}
R_{ex}=R_{ref} \frac{I^{meas}(Q)-I^{meas}_{solv}}{I^{meas}_{ref}} \left( \frac{n}{n_{ref}} \right)^{2}
\label{IQ}\
\end{equation}
where the subscripts ``solv'' and ``ref '' refer to the solvent and to the standard reference of toluene. Q is the exchanged wave vector, R$_{ref}$ the Rayleigh ratio of the reference at $\lambda$=632.8 nm, n and n$_{ref}$ the refractive index of the sample and of the reference respectively.
\\
The excess Rayleigh ratio R$_{ex}$ can be also expressed as the product of different terms:
\begin{equation}
R_{ex}=K c M_{w} P(Q) S(Q)
\label{IQ2}\
\end{equation}
where K=$\frac{4 \pi^{2} n^{2}}{N_{A} \lambda^{4}} \left( \frac{dn}{dc} \right)^{2}$ is the optical constant, c the mass concentration, M$_{w}$ the molecular weight, P(Q) and S(Q) are the normalized form factor and structure factor respectively.
\\
dn/dc is the refractive index increment representing the difference in refractive index between the sample and the solvent and was obtained from measurements of IPN dispersions as a function of concentration with an Abbe refractometer in a range close to that used in SLS experiments. In dilute condition (small concentrations c) and in the low Q limit, the structure factor can be approximated with its value at zero Q, S(0) = (1+ 2A$_{2}$M$_{w}$ c)$^{-1}$ , and the form factor  with P(Q)=(1 + Q$^{2}$ R$_{g}^{2}$/3)$^{-1}$ where A$_{2}$ is the second virial coefficient which describes the solute-solvent interactions and R$_{g}$ the gyration radius. Combining equation (\ref{IQ}) and equation (\ref{IQ2}) the Zimm equation is obtained:
\begin{equation}
\frac{Kc}{R_{ex}}=\frac{1}{M_{w}} \left( 1+ \frac{Q^{2}R_{g}^{2}}{3}+O(Q^{4}) \right)+ 2A_{2}c+ O(c^{2}).
\label{ZimmPlot}\
\end{equation}
By plotting Kc/R$_{ex}$  as a function of Q$^{2}$+k$'$c, with k$'$ a constant chosen arbitrarily, one can obtain the Zimm plot from which molecular weight, radius of gyration and second Virial coefficient A$_{2}$ can be determined. The intercept of a linear interpolation at fixed Q, therefore, yields an inverse molecular weight, while the slope is proportional to the second virial coefficient. Similarly, the radius of gyration is obtained by linear interpolation of the Zimm plot at a fixed concentration. More details on the method are reported in a previous paper \cite{MicaliCPC2018} of some of the authors of this.
\subsection{Determination of microgel generalized volume fraction $\zeta$ at different temperatures}
Microgels are characterized by a volume phase transitions marked by a shrinkage of the particles with a consequent change in volume and volume fraction. This determines also a strong temperature dependence of the \textit{k}-values. 
In order to obtain the volume fraction as a function of temperature, once determined $\zeta$ at room temperature from viscosity measurements, one can exploit the concept that for isotropic swelling  $\zeta$ can be related to the particle size R as \cite{SenffJCP1999, CarrierJR2009}:
\begin{equation}
\zeta(T)=\zeta(T_{0}) \cdot \left( \frac{R(T)}{R(T_{0})} \right) ^{3}
\label{eqCarrier}
\end{equation}
where $R(T_0)$ is the particle diameter in the reference state and $R(T)$ is the particle diameter at a given temperature both measured through DLS.
%
%
%%%%%%%%%%%%%%%% Section: Experimental method %%%%%%%%%%
\section{Experimental method}

\subsection{\textit{Materials}}

N-isopropylacrylamide (NIPAM) monomer (Sigma-Aldrich) and N,N'-methylene-bis-acrylamide (BIS), (from Eastman Kodak) were recrystallized from hexane and methanol respectively, dried under reduced pressure (0.01$\unit{mmHg}$) at room temperature and stored at $253\unit{K}$.
Acrylic acid (AAc) monomer (Sigma-Aldrich) was purified by distillation (40$\unit{mmHg}$, 337$\unit{K}$) under nitrogen atmosphere in presence of hydroquinone and stored at 253$\unit{K}$.
Sodium dodecyl sulphate (SDS), 98\% purity, surfactant potassium persulfate (KPS), 98\% purity and ammonium persulfate (APS), 98\% purity, N,N,N',N'-tetramethylethylenediamine (TEMED), 99\% purity, ethylenediaminetetraacetic acid (EDTA) (a chelating agent for purifying dialysis membranes), and NaHCO$_{3}$, (all purchased from Sigma-Aldrich) were used as received.
Ultrapure water (resistivity: 18.2$\unit{M\Omega/cm}$ at 298$\unit{K}$) was obtained with Arium\textregistered$\:$ %%%%%%%%Come togliere corsivo dalla formula%%%%%%
pro Ultrapure water purification Systems, Sartorius Stedim.
All other solvents were RP grade (Carlo Erba) and were used as received.
A dialysis tubing cellulose membrane, MWCO 14000$\unit{Da}$, (Sigma-Aldrich) was cleaned before use by washing with running distilled water for 3$\unit{h}$, treated at 343$\unit{K}$ for 10$\unit{min}$ into a solution containing 3.0\% NaHCO$_{3}$ and 0.4\% EDTA weight concentration, rinsed in distilled water at 343$\unit{K}$ for 10$\unit{min}$ and finally in fresh distilled water at room temperature for 2$\unit{h}$.

\subsection{\textit{Microgel synthesis}}

\begin{table*}[t]
\centering
\caption{Characteristic parameters of IPN microgels at four different PAAc contents. $k$ is the conversion parameter from concentration to volume fraction derived as described in the text, $R$ is the particle radius measured through dynamic light scattering at T=298$\unit{K}$ and T=311$\unit{K}$ at weight concentration C$_{w}$=0.01$\%$.
C$_{PAAc}$, C$_{PNIPAM}$ and C$_{BIS}$ are the contents of PAAc, PNIPAM and crosslinker BIS within particles measured through  elemental and $^{1}$H-NMR analysis \cite{MicaliCPC2018}.}
\begin{tabular}{ccccccc}
\toprule
C$_{\mathrm{PAAc}} \, (\%)$  & $k_{/100}$ & R$(298K) \, (\mathrm{nm})$ & R$(311K) \, (\mathrm{nm})$ & C$_\mathrm{PNIPAM} \ (\%)$ & C$_\mathrm{BIS} \, (\%)$ & C$_{BIS}$/C$_{PAAc}$ \\ 
\midrule
%2.6 &  $0.37\pm...$ & / & / & 95.44 & 2.98 & 1.15\\
10.6 &  $0.72\pm0.03$ & $89\pm2$ & $52\pm1$ & 88.3 & 1.1 & 0.10\\
19.2 &  $0.78\pm0.01$ & $126\pm3$ & $91\pm2$ & 73.6 & 7.2 & 0.38\\
24.6 &  $2.33\pm 0.02$ & $159\pm3$ & $143\pm3$ & 67.7 & 7.7 & 0.31\\
28.0 &  $2.87\pm0.11$ & $252\pm14$ & $240\pm18$ & 62.2 & 9.8 & 0.35\\
\bottomrule
\label{table1}
\end{tabular}
\end{table*}
First poly(N-isopropylacrylamide) microgels were synthetized by a precipitation polymerization method following the procedure described by Pelton et al. \cite{PeltonColloids1986}.
% la ref nella tua bibliografia era {Pelton_Campione}
Then IPN microgels were obtained interpenetrating PAAc network in PNIPAM microgels by a sequential free radical polymerization method \cite{XiaLangmuir2004, HuAdvMater2004}.
%nella tua bibliografia si chiamavano: \cite{XiaLan}\cite{HuGelation}?
In particular (24.162$\pm$0.001)$\unit{g}$ of NIPAM monomer, (0.4480$\pm$0.0001)$\unit{g}$ of BIS and (3.5190$\pm$0.0001)$\unit{g}$ of SDS surfactant were solubilized in 1560$\unit{mL}$ of ultrapure water and transferred into a 2000$\unit{mL}$ five-necked jacked reactor equipped with condenser and mechanical stirrer.
The solution, heated at (343$\pm$1)$\unit{K}$, was deoxygenated by purging with nitrogen for 1$\unit{h}$.
Then the polymerization was initiated by adding (1.0376$\pm$0.0001)$\unit{g}$ of KPS, dissolved in 20$\unit{mL}$ of deoxygenated water and carried out for 16$\unit{h}$.
After this time, the resultant PNIPAM microgel was purified by dialysis against distilled water for two weeks changing water frequently.
The final weight concentration C$_{w}$ of PNIPAM micro-particles was C$_{w}$=1.06$\%$ as determined by gravimetric measurements.
In the second step (140.08$\pm$0.01)$\unit{g}$ (C$_{w}$=1.06$\%$) of the recovered PNIPAM dispersion was diluted with ultrapure water up to a volume of 1260$\unit{mL}$ into a 2000$\unit{mL}$ five-necked jacketed reactor, kept at (295$\pm$1)$\unit{K}$ by circulating water.
Then 5 mL of AAc monomer and (1.1080$\pm$0.0001)$\unit{g}$ of BIS crosslinker were added to the dispersion and the mixture was deoxygenated by bubbling nitrogen inside for 1$\unit{h}$.
0.56$\unit{mL}$ of accelerator TEMED were added and the polymerization was started with (0.4447$\pm$0.0001)$\unit{g}$ of initiator APS.
\\
IPN microgels with four different PAAc contents were obtained by stopping the polymerization at different reaction times by exposing to air.
They were purified by dialysing against distilled water with frequent water changes for two weeks, and then lyophilized and redispersed in water to form a sample at weight concentration C$_{w}=1.0\%$.
The PAAc weight concentration (C$_{PAAc}$) of the four IPN samples were determined by combination of elemental and $^{1}$H-NMR analysis as described in reference \cite{MicaliCPC2018}, and they are  C$_{PAAc}$=10.6$\%$, C$_{PAAc}$=19.2$\%$, C$_{PAAc}$=24.6$\%$ and C$_{PAAc}$=28.0$\%$.
Despite the different PAAc content the pKa of microgels should vary slowly as demonstrated in a recent work \cite{SwiftSM2016} where the pH-responsive behaviour of polyacrylic acid chains with different molar mass was studied. The pKa obtained from the titration data varied in a narrow range around 4.5 is in good agreement with previous works \cite{AnghelPoly1998, ArnoldJCS1957}.
Each PAAc content took almost one month to be synthesized.
Samples at  weight concentrations C$_{w}$ $<$ 1.0$\%$ were obtained by dilution with distilled water from the same stock suspension at C$_{w}$=1.0$\%$ and adjusted at pH=5.5 while samples at C$_{w}$ $>$ 1.0$\%$ were obtained by redispersing a sample lyophilized at C$_{w}$=3.0$\%$.
%
%
%\sil{\textbf{Non sappiamo se metterlo o meno: The dilution from stock solutions at C$_{w}$ = 1.0$\%$ does not affect significantly the final pH and therefore we can consider as if the pH of the solutions was kept constant for the determination of the volume fraction.}}
%
%
%
%
\subsection{\textit{Dynamic Light scattering measurements}}
Particle size of IPN microgels have been determined through dynamic light scattering (DLS) measurements using an optical setup based on a solid state laser (100$\unit{mW}$)  with monochromatic ($\lambda$ = 642$\unit{nm}$) and polarized beam. Measurements have been performed at a scattering angle $\theta=90^{\circ}$ corresponding to a scattering vector Q=0.018 $\unit{nm^{-1}}$ according to the relation $Q = (4\pi n/ \lambda)sin(\theta/2)$. The average size of the particles has been obtained through the Stokes-Einstein relation $R = K_BT/6\pi\eta Dt$ where $K_B$ is the Boltzmann constant, $\eta$ the viscosity and $D_t$ the translational diffusion coefficient related to the relaxation time $\tau$ through the relation: $\tau = 1/(Q^2 D_t)$. 
The relaxation time $\tau$ was obtained fitting the autocorrelation function of scattered intensity through the Kohlrausch-William-Watts expression, $g_2 (Q,t)=1+b[(\exp{-t/\tau})^{\beta}]^2$, with the stretching exponent $\beta$ .
Measurements have been performed in dilute conditions at T=298$\unit{K}$ and T=311$\unit{K}$ in the swollen and shrunken state respectively and the radii of the microgels are reported in table \ref{table1}.

%%%%%%%%%%%%%%%Rheological characterization %%%%%%%%%%%%%%%%%%

\subsection{\textit{Rheological measurements}}

Rheological measurements have been performed with a stress-controlled Discovery Hybrid Rheometer-1 from TA instruments (New Castle, DE, USA)  by using a cone-plate geometry (plate diameter=40$\unit{mm}$, cone angle=1$^{\circ}$, truncation gap=0.027$\unit{mm}$) and with a rotational rheometer Anton Paar MCR102 with a cone-plate geometry (plate diameter=49.97 mm, cone angle=2.006$^{\circ}$, truncation=212 $\mu$m) both equipped with a Peltier temperature controlling unit.
Flow curves were determined in the shear rate range (10$^{0}$-10$^{4}$) $\unit{s^{-1}}$ at 298$\unit{K}$ and 311$\unit{K}$. A stepwise increase of the stress was applied with an equilibration time of 30$\unit{s}$. Four different PAAc concentrations were investigated in the weight concentration range C$_w$ = (0.01-1.5)\%.

\section{Results and discussion}

%
%
%%%%%%%%%%Figura%%%%%%%%%% fig1   (1)
\begin{figure}[t]
\centering
\includegraphics[clip, scale=0.54]{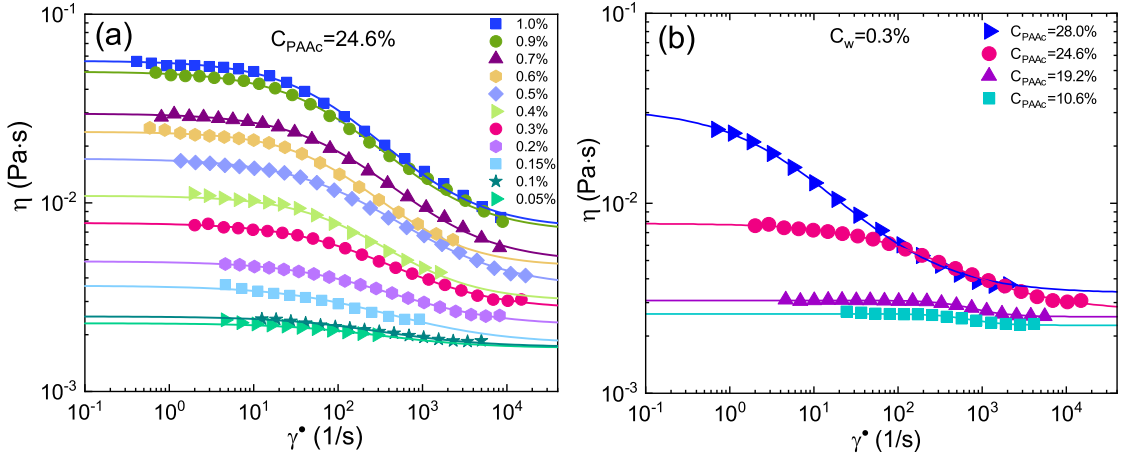}
\caption{Shear viscosity $\eta$ as a function of shear rate $\dot{\gamma}$ at T=298$\unit{K}$ (a) in the weight concentration range (0.05-1.0)$\%$ for IPN microgels at C$_{PAAc}$=24.6$\%$ and (b) at weight concentration  0.3\% for IPN microgels at four PAAc contents. Lines are fits according to the Cross model (equation (\ref{cross_model})).}
\label{fig1}
\end{figure}
In this section we report the main results of this experimental work.
Firstly, in order to determine the volume fraction $\zeta$ of IPN microgels from viscosity measurements, we performed flow curves at four PAAc contents for different concentrations C$_{w}$. 
In Figure \ref{fig1}(a) the viscosity as a function of shear rate $\dot{\gamma}$ is reported for IPN microgels at C$_{PAAc}$=24.6$\%$ as an example and the flow curves display a shear thinning behaviour that is more evident at high weight concentrations.
Data are fitted through the Cross model (equation (\ref{cross_model})) that allows to obtain  the zero-shear viscosity $\eta_{0}$. 
%
%The flow curves in figure \ref{fig1}(a) show a shear thinning behaviour that is more evident at high weight concentrations.
In Figure \ref{fig1}(b) the comparison among four different PAAc contents,  C$_{PAAc}$=10.6$\%$, C$_{PAAc}$=19.2$\%$, C$_{PAAc}$=24.6$\%$ and C$_{PAAc}$=28.0$\%$, at fixed weight concentration (C$_w$=0.3$\%$) shows a clear increase of the viscosity with increasing the PAAc content of microgel particles.

\begin{figure*}[h!]
\centering
\includegraphics[trim=50 0.8cm 50 25, clip, scale=0.35]{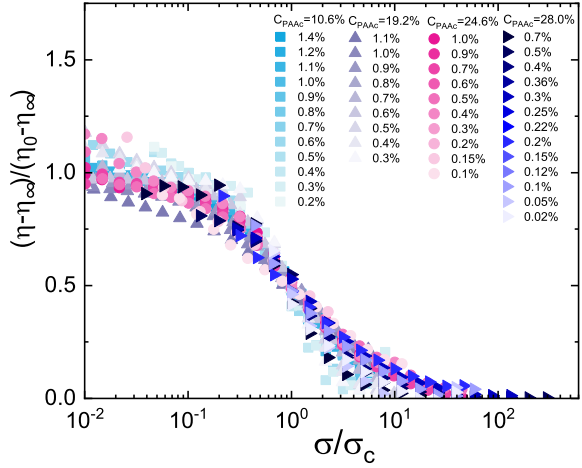}
\caption{Scaled plot of $(\eta-\eta_{\infty})/(\eta_{0}-\eta_{\infty})$ against reduced shear stress for IPN microgels at four PAAc contents (C$_{PAAc}$=10.6$\%$, C$_{PAAc}$=19.2$\%$, C$_{PAAc}$=28.0$\%$ and C$_{PAAc}$=24.6$\%$), at different weight concentrations and at T=298K. All data follow a unique mastercurve. }
\label{fig2}
\end{figure*}

\begin{figure}
\centering
\includegraphics[trim=0.5cm 1cm 12cm 0.4cm, clip, scale=0.32]{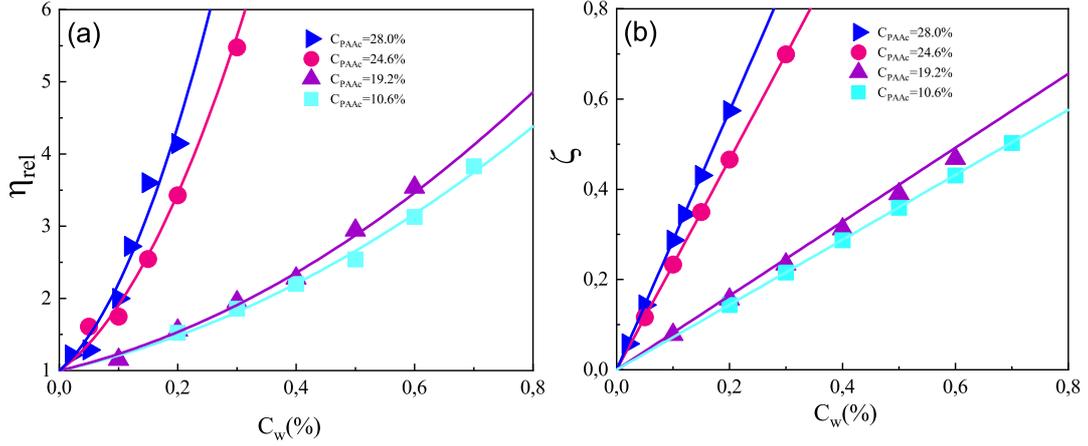}
\caption{(a) Relative viscosity from equation (\ref{etarel}) as a function of weight concentration at T=298K. Data are fitted according to Einstein-Batchelor model of equation (\ref{EB}). (b) Volume fraction $\zeta$ versus weight concentration. The angular coefficient of the linear fit represents the conversion constant k between weight concentration C$_{w}$ and volume fraction $\zeta$ as reported in equation (\ref{costante_k}).}
\label{fig3}
\end{figure}

Later on, after having obtained the characteristic fit parameters for all flow curves at all different C$_{PAAc}$ and C$_{w}$, the normalized shear viscosity $(\eta-\eta_{\infty})/(\eta_{0}-\eta_{\infty})$ is reported in Figure \ref{fig2} as a function of $\sigma/\sigma_c$ \cite{SenffCPS2000, deKruifJCP1985} where $\sigma$ is the shear stress and $\sigma_c$ the critical shear stress defined as the stress corresponding to the intermediate shear viscosity $\eta=(\eta_{0}-\eta_{\infty})/2$. All data for different investigated C$_{PAAc}$ and C$_w$ fall into a single mastercurve.
%
%
%%%%%%%%%%%Figura%%%%%%%%%% fig2
The relative viscosity $\eta_{rel}=\eta_0/\eta_s$ (\ref{etarel}) is plotted  in Figure \ref{fig3}(a) as a function of weight concentration in dilute regime and a fit through the Einstein-Batchelor expression (equation \ref{EBCw}), letting $\eta_s$ as a free parameter (see SI), provides the conversion factor $k$ (equation (\ref{costante_k})) that directly links the polymer concentration C$_{w}$ to the volume fraction $\zeta$.
Moreover, Figure \ref{fig3}(a) shows a strong C$_{PAAc}$ dependence of the relative viscosity  with a more pronounced growth for the samples at  C$_{PAAc}$=24.6$\%$ and C$_{PAAc}$=28.0$\%$  confirming that the viscosity is highly influenced by polyacrylic acid and weight concentration.
The volume fraction $\zeta$ as a function of concentration C$_{w}$ is shown in Figure \ref{fig3}(b) for the four PAAc contents investigated and the $k$ parameter is the slope of the curves.
Values of the $k$ parameter are reported in table \ref{table1} for each C$_{PAAc}$ at room temperature, showing that $k$ becomes greater with increasing microgel radius and therefore polyacrylic acid and crosslinker content.
Interestingly, these results are in agreement with findings reported for PNIPAM in reference \cite{SenffJCP1999} where an increase of $k$ is associated to an increase of particle radius, in this case tuned by temperature.
Moreover, high polyacrylic acid contents also determine a reduction of particles softness \cite{NigroJCIS2019}, implying that $k$ and hence $\zeta$ increase with decreasing particle softness.
Therefore, for IPN microgels, the constant $k$ and hence the volume fraction, are strictly linked to the particle softness controlled by polyacrylic acid content and crosslinker density during synthesis.
This finding is also in agreement with a recent work on PNIPAM \cite{ScottiSM2020}, where the relative viscosity plotted versus the generalized volume fraction $\zeta$ shows that the  onset of the divergence of viscosity is shifted to higher values of $\zeta$ for PNIPAM samples with lower crosslinker content and therefore higher softness.
In summary, particles with higher polyacrylic acid content are characterized by larger size, greater $k$ and larger volume fractions when compared to less soft particles.
%
%
%%%%%%%%%%Figura%%%%%%%%%% fig3   
%%%%%%%%%%%%%%%%%%%%%%%%%
%
%

\begin{figure}[h!]
\centering
\includegraphics[clip, scale=0.35]{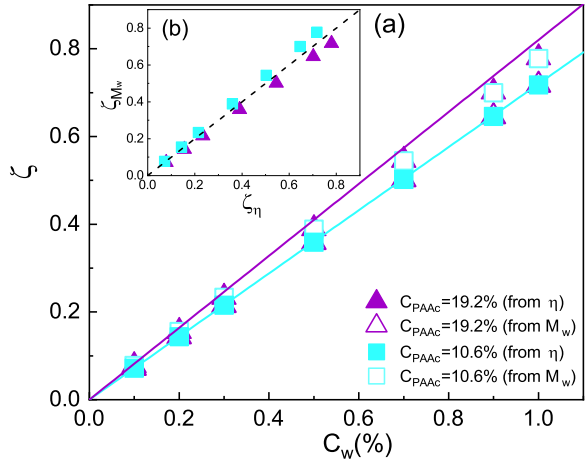}
\caption{Volume fraction versus weight concentration for IPN microgels with C$_{PAAc}$=10.6$\%$ and C$_{PAAc}$=19.2$\%$, at T=298K, as obtained by rheology (full symbols) and SLS measurements (open symbols). Lines are linear fit to the data. (b) Volume fraction obtained throug SLS from the molecular weight, $\zeta_{M_{w}}$ as a function of the volume fraction found by rheology, $\zeta_{\eta}$. The Straightline has slope 1.}
\label{fig4}
\end{figure}

With the aim of validating the volume fractions obtained through rheological flow curves, in Figure \ref{fig4}(a), a comparison between the packing fraction values obtained through rheology and SLS measurements \cite{MicaliCPC2018} is reported for two IPN microgels (C$_{PAAc}$=10.6$\%$ and C$_{PAAc}$=19.2$\%$).

The values of $\zeta$  from SLS measurements have been calculated through equation (\ref{zetavolume}) where R$_0$ and M$_w$ are those reported in reference \cite{MicaliCPC2018}. The volume fraction obtained from rheology, $\zeta_{\eta}$, is in good agreement with the one found from the molecular weight, $\zeta_{M_w}$. 
In order to better visualize the comparison between the two methods, $\zeta_{M_w}$ is plotted versus $\zeta_{\eta}$ in Figure \ref{fig4}(b) and we can confirm that, within the experimental error, the curves are close to that with slope 1, corresponding to perfect agreement.

\begin{figure}[h!]
\centering
\includegraphics[clip, scale=0.36]{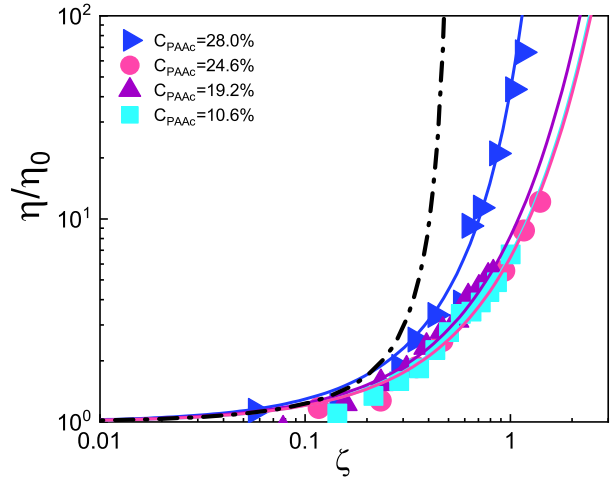}
\caption{Normalized shear viscosity as a function of volume fraction $\zeta$ at T=298K. Lines represent fits through Arrhenius model
(equation (\ref{Arrhenius})) for IPN microgels with  C$_{PAAc}$=10.6$\%$,  C$_{PAAc}$=19.2$\%$ and C$_{PAAc}$=24.6$\%$, Vogel-Fulcher-Tamman model (equation (\ref{VFT})) for IPN microgel with C$_{PAAc}$=28.0$\%$ and Vogel-Fulcher-Tamman for hard sphere with A=1 and $\zeta_{0}$=0.58 (dash-dot line)}
\label{fig5}
\end{figure}

The normalized viscosity behaviour at all the investigated concentrations, in addition to those in the dilute regime, is reported in Figure \ref{fig5} at different PAAc (and crosslinker) contents.
%
%
%%%%%%%%%%Figura%%%%%%%%%% fig5   

To avoid misunderstandings, we define the viscosity at zero shear rate $\eta_{0}$ obtained from the Cross model as $\eta$, while $\eta_{0}$ represents the viscosity in the limit of zero volume fraction from equations (\ref{Arrhenius}) and (\ref{VFT}).
It can be seen that the growth of the viscosity is more gradual for lower PAAc (and crosslinker) contents and it is well fitted with an Arrhenius dependence (equation \ref{Arrhenius}). In contrast, at the highest investigated PAAc (28.0$\%$) where a Vogel-Fulcher-Tamman dependence (equation (\ref{VFT})) well reproduces the data, its divergence  moves to lower volume fractions and its increase is much more steep. Data are also compared with the behaviour of  hard
spheres viscosity plotted as a dash-dot line according to equation (\ref{VFT}) with A=1 and $\zeta_{0}$=0.58.
These outcomes also evidence that, with decreasing particles softness, the viscosity divergence occurs at lower $\zeta$  up to the limiting value of hard spheres, $\zeta_{0}$=0.58, demostrating how the peculiarity of microgels to swell, deform and interpenetrate allows $\zeta$ to reach, at high concentrations, values above 1 for softer particles. 

\begin{figure}[t]
\centering
\includegraphics[clip, scale=0.35]{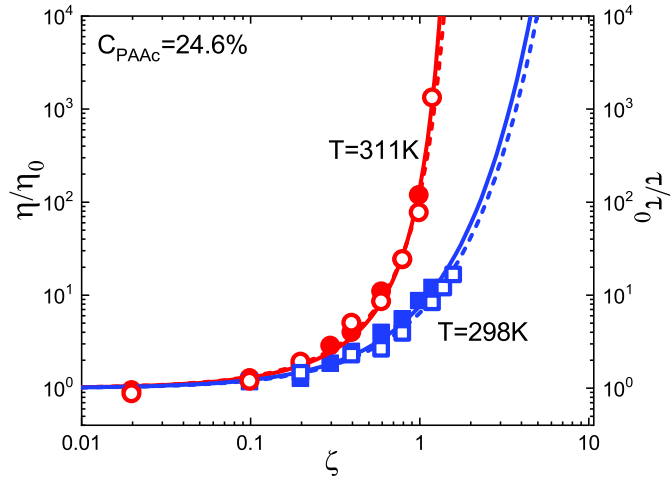}
\caption{Normalized shear viscosity (closed symbols) and relaxation time (open symbols) vs volume fraction below (T=298$\unit{K}$) and above (T=311$\unit{K}$) the volume phase transition for IPN microgel with C$_{PAAc}$=24.6$\%$.
Lines represent fits according to the Arrhenius expression (equation (\ref{Arrhenius})) for T=298$\unit{K}$ and to the VFT expression (equation (\ref{VFT})) for T=311$\unit{K}$.}
\label{fig6}
\end{figure}

These findings are in good agreement with the behaviour of the normalized structural relaxation time $\tau/\tau_0$ reported in a previous work \cite{NigroMacromol2020} for IPN microgels with different PAAc contents and at a temperture, T=311$\unit{K}$, above the volume phase transition. Here $\tau/\tau_0$ increases more rapidly at higher PAAc contents with increasing weight concentration following always a VFT behaviour (equation \ref{VFT}).
%%%%%%%%%%Figura%%%%%%%%%% fig4  
%
%
Accordingly, a direct comparison between normalized structural relaxation time and viscosity for an IPN microgel at C$_{PAAc}$=24.6$\%$ is reported in Figure \ref{fig6} at two temperatures, T=298$\unit{K}$ and T=311$\unit{K}$, below and above the volume phase transition in the swollen and shrunken state respectively.
In order to calculate the volume fraction at T=311$\unit{K}$,  viscosity measurements as a function of concentration in dilute regime were performed and the conversion factor $k$ between weight concentration and volume fraction was found through the Batchelor model as performed at room temperature (T=298$\unit{K}$). Moreover exploting the concept that for isotropic swelling, $\zeta$ can be related to the particle size R through equation (\ref{eqCarrier}), the comparison
between the volume fraction  measured at T=298$\unit{K}$ and that calculated at T=311$\unit{K}$ is reported in SI. 
We found that structural relaxation time and viscosity are in excellent agreement, both growing gradually and being well described by an Arrhenius dependence at T=298$\unit{K}$ below the VPT while they increase more steeply with at T=311$\unit{K}$ above the VPT following a Vogel-Fulcher-Tammann behaviour (equation \ref{VFT}). Also in this case, two different models are used below and above the VPT since the microgel softness increases with decreasing temperature across the volume phase transition.
Therefore, it is clear that the volume fraction dependence of viscosity and relaxation time is well described by an Arrhenius law for softer particles as in the case of low PAAc (and crosslinker) at temperatures below the VPT (Figure \ref{fig5} and \ref{fig6}). On the contrary, they follow a VFT law for less soft particles at high PAAc (C$_{PAAc}$=28.0$\%$) and crosslinker content (Figure \ref{fig5}) or at temperatures above the VPT (Figure \ref{fig6}), validating the idea  that softness can be controlled by PAAc, crosslinker and temperature.
%
%
%
%
%%%%%%%%%%Figura%%%%%%%%%% fig6 
%%begin{figure}[t]
%\centering
%\includegraphics[clip, scale=0.38]{fig6.eps}
%\includegraphics[trim=40 1cm 100 50, clip, scale=0.38]{fig6.eps}
%\caption{Normalized shear viscosity (closed symbols) and relaxation time (open symbols) vs volume fraction below (T=298$\unit{K}$) and above (T=311$\unit{K}$) the volume phase transition for IPN microgel with C$_{PAAc}$=24.6$\%$.
%Lines represent fits according to the Arrhenius expression (equation (\ref{Arrhenius})) for T=298$\unit{K}$ and to the VFT expression (equation (\ref{VFT})) for T=311$\unit{K}$.}
%\label{fig6}
%\end{figure}
%
%
\section{Conclusions}
In this work we have determined the volume fraction of an interpenetrated polymer network  microgel composed of poly(N-isopropylacrylamide) and polyacrylic acid through viscosity measurements at different PAAc contents and weight concentrations. Experimental flow curves have been fitted through the Cross model that allows to gain information on the zero-shear viscosity. This is used both to plot a mastercurve of viscosity data and to determine, through the Einstein-Batchelor equation, the shift factor $k$ to convert the weight concentration to volume fraction. 
Moreover viscosity behaviour as a function of $\zeta$ has highlighted an Arrhenius-like and a Vogel-Fulcher-Tammann dependence for microgels with low and high PAAc content respectively. Interestingly a comparison with hard spheres indicates a steepest increase of the viscosity with decreasing particles softness. A similar scenario is found below and above the VPT, in good agreement with the structural relaxation time behaviour, indicating that softness in IPN microgels can be tuned with PAAc, crosslinker and temperature and that, depending on particle softness, two different routes are followed.

\addcontentsline{toc}{chapter}{Bibliography}
%\begin{thebibliography}{10}
\bibliography{bibliografia}
\bibliographystyle{unsrt}	
%\end{thebibliography}

\end{document}